\documentclass[prl,twocolumn]{revtex4}
\usepackage{graphicx,color,epstopdf}
\usepackage{amsmath,amsfonts,enumerate,amsthm,amssymb,mathtools}
\usepackage{enumitem}
\usepackage{thmtools,thm-restate}
\usepackage{bbold}
\usepackage{hyperref}
\usepackage{multirow}
\usepackage{subfigure} 
\usepackage{mathdots} 

\hypersetup{
linkcolor=blue, 
citecolor=blue 
}

\usepackage{hyperref}

\begin{document}

\draft

\title{Zero knowledge convincing protocol on quantum bit
is impossible}

\author{Pawe\l{} Horodecki $^1$, Micha\l{} Horodecki $^2$, Ryszard Horodecki $^1$}

\address{
$^1$ Faculty of Applied Physics and Mathematics, National Quantum Information Centre, Gda\'nsk University of Technology, \\
80--233 Gda\'nsk, Poland \\
$^2$ Institute of Theoretical Physics and Astrophysics, National Quantum Information Centre, Faculty of Mathematics, 
Physics and Informatics University of Gda\'nsk, 80-308 Gda\'nsk, Poland \\
}

\begin{abstract}
It is one of fundamental features of quantum formalism that on
one hand it provides a new information processing
resources and on the other hand puts fundamental constraints
on the processing of quantum information implying
``no-go'' theorems for cloning \cite{noclon1,noclon2,noclon3},
 bit commitment \cite{bc1,bc2} and deleting \cite{nodel}
in quantum theory.
Here we ask about possibility of ``zero knowledge''
scenario which, for its simplicity, can be considered as a
quantum primitive or model scenario for any problems of similar kind.
Consider two parties: Alice and Bob and suppose that
Bob is given a qubit system in a quantum state $\phi$,
unknown to him. Alice knows $\phi$ and she is supposed to convince Bob that she knows
$\phi$ sending some test message. Is it possible for her to convince
Bob providing him "zero knowledge" i. e. no information about $\phi$
he has? We prove that there is no "zero knowledge" protocol of
that kind. In fact it turns out that
basing on Alice message, Bob (or third party - Eve -
who can intercept the message) can synthetize a copy of the unknown
qubit state $\phi$ with nonzero probability. This "no-go" result puts general
constrains on information processing where information {\it about} quantum
state is involved.
\end{abstract}

\maketitle

Consider first the most general test message from Alice.
It can involve some {\it classical information}
(some data, function encoded in  classical bits)
as well as purely {\it quantum information} represented by
quantum register or, in other words, quantum
system in some state. In her message she can, for example
order to perform quantum computing of some problem and foresee
the result or even - in general - she can order Bob
to run	both quantum and classical Turing machines to check
some of her predictions. She must make	some predictions
however, as the message is supposed to {\it test} her knowledge.
Thus, in general, Bob must perform some {\it measurement} to check
her predictions.

{\it The general form of test message within quantum formalism .-}
All the above can taken into account in the
{\it test message} (or test in brief) sent
form Alice to Bob consisting of three elements
 (i) {\it classical} prescription of some quantum
operation, (ii) possibly -  some ancilla in {\it quantum} state
Alice prepared together with (iii) result of the operation.
The quantum operation is supposed to act in general on both
ancilla and Bob's state, but this action could be in particular
trivial i. e. not affecting some of them at all.
Note that all classical information (say classical bits as
classical Turing machine etc.) can be included in the description
of quantum operation. In fact it can be states of ancillas prepared by Bob.
This is because Bob is supposed to perform finally the measurement
in which his qubit as well as Alice ancilla are, in general,
supposed to subject. All the results of Bob operation (even
if there is more than one) predicted by Alice can be included
in 
the output of a general quantum measurement.

{\it The conditions for convincing test messages .-}
Let us now consider  what it would mean to convince Bob in the
above scenario or, in other words, which test message from Alice
to Bob is {\it convincing}?
It is obvious that if Alice knows the state
and she wants to convince Bob about it then
{\it the result of the operation she predicts must occur
with probability one i. e.  with certainty}.
But it is not all: Alice could try to cheat proposing the operation
which would give some result with certainty
independently on her knowledge about $\phi$.
For example Alice could order Bob to prepare
the spin-$\frac{1}{2}$ in state ``up''
$| \uparrow_{\mathop{\bf \hat{z}}} \rangle$
and predict that if he measures the spin component
of the particle along $\mathop{\bf \hat{z}}$ then he will
get result ``up''. Of course, Alice prediction is right with
probability one but has nothing to do with
her knowledge about $\phi$ at all.
To avoid this the convincing test message should have the property
that if Alice does not know $\phi$ then there is nonzero
probability that the result she gives (as prediction)
to Bob will not occur i. e.  there is nonzero probability of
revealing that she cheats. We can summarise the above
in the following

 {\it Condition 1 .- Any convincing test message
from Alice to Bob should have the property
that Alice's prediction occurs with certainty
if and only if she knows the state $\phi$.}

There is, however, a problem that Alice can
send to Bob the convincing test message, but he may not be
able to check whether the condition 1 is satisfied
i. e. whether the result of the test
is not independent on Alice knowledge about $\phi$.
On the other hand checking whether the test
satisfies the condition he could
be forced to destroy some quantum information
(which can not be cloned) about the test
and will not be able to carry out the test
(or any test equivalent to it). To avoid those two
possibilities of that kind we shall postulate
the natural condition

{\it Condition 2 .-
For any  convincing test message Bob must be able to check
the condition 1 for the message in such a way
that he still can find out whether the test itself works.}

Any test message satisfying conditions 1 and 2 we shall call
{\it convincing}.
First we shall prove that convincing messages exist. Consider
the following protocol.
Alice sends Bob only classical message (with no ancilla):
``Please measure the spin value along the axis $\mathop{\bf \hat{n}}$.
You will certainly get result ``up''.  This corresponds
 to the state you have''. (where Alice's state $|\phi\rangle$ is ``up'' eigenvector along $\mathop{\bf \hat{n}}$ axis).
From the above Bob can see himself that the protocol
satisfies the condition
1 as if Alice does not know the state exactly (up to
some phase factor) then from her point of
view it is random with some nontrivial probability distribution.
So it is likely that $\phi$ has both ``up'' and ``down'' components
nonzero along $\mathop{\bf \hat{n}}$ i. e. that $\phi=\alpha|
\uparrow_{\mathop{\bf \hat{n}}} \rangle
+ \beta |\downarrow_{\mathop{\bf \hat{n}}} \rangle$ with $\beta\neq 0$.
Then there is nonzero probability of
result ``down'' contrary to what Alice predicted.
The above protocol is based on ``convincing'' test.
Obviously it is not zero knowledge protocol
as the full information about $|\phi\rangle$
has been transferred from Alice to Bob.

{\it Fully classical protocols .-}
Below we shall focus on the protocols, that we call fully classical
ones in which there is only classical information transfer from Alice to Bob.
We shall briefly prove the following observation

{\it Observation .- To make the convincing test nontrivial
i. e. not carrying all information about $\phi$
the transfer of quantum information (represented by
ancilla) form Alice to Bob is necessary.
In other words fully classical protocols are completely
trivial from ``zero knowledge'' point of view.}

As we discussed before, the whole classical part of message can be
included in the classical description of some quantum operation
and its result which Alice predicts. The most general quantum operation Bob
can perform on the state is the generalised quantum
measurement mathematically represented by completely positive map.
According to general results of quantum measurement theory
it can be written as follows
\begin{equation}
|\phi \rangle \langle \phi| \rightarrow
\varrho=\sum_{i}V_{i} |\phi \rangle \langle \phi| V_{i}^{\dagger}
\end{equation}
where $\sum_{i}V_{i}^{\dagger}V_{i}$ is equal to $2 \times 2$
identity matrix. The indices $i$ correspond to elementary results
of the measurement. The most general result predicted
by Alice can be that  Bob's result $i_{0}$ will belong to some
subset of indices $I$. Now we ask about the information which is
carried in Alice test message if she does not cheat.
Then her prediction must occur with probability one.
In quantum measurement theory it means that
\begin{equation}
p=Tr(\sum_{i\in I} V_{i}^{\dagger} V_{i}|\phi\rangle \langle \phi|)=1
\label{jeden}
\end{equation}
On the other hand, the measurement theory asserts that the hermitian operator
$X=\sum_{i \in I} V_{i}^{\dagger}V_{i}$ has eigenvalues $\lambda$
satisfying $0 \leq \lambda \leq 1$.
So (\ref{jeden}) means that $|\phi\rangle$ is an eigenvector of $X$.
We have, however, one more condition for test message to be convincing:
if Alice does not know the state then there must be nonzero probability
of revealing it.
This means that in such case the result she predicted
{\it should occur with probability strictly less then one}.
This  can be very simply expressed as
\begin{equation}
p'=Tr(\sum_{i\in I}V_{i}^{\dagger}V_{i}|\phi'\rangle \langle \phi'|)<1
\ \ \mbox{for}\ \ \mbox{any} \ \ \phi' \neq \phi.
\label{jeden'}
\end{equation}
But, following the spectral property of $X$ it means that the
state $|\phi\rangle$ is the only eigenvector of $X$
corresponding to the eigenvalue $\lambda=1$.
Bob (or Eve) does not know the state. But he is
given description of $V_{i}$-s and the set of indices on the paper
(or, say, computer disc). He can
now calculate $X$, diagonalise,  find the unique
eigenvector corresponding to unit eigenvalue - the vector is nothing
but $\phi$. He can finally perform the physical measurement
of the observable $X$ on his particle to test
whether the Alice message protocol works (i. e.  whether
she does not cheat).
Let us summarise. If Alice knows $\phi$ and wants to convince Bob about it
sending only classical information, then any test message
form her must contain {\it full} information about
the Bob's state $\phi$. Bob is able to
get the information and still check whether her test operation works.
The crucial observation here is that given the operation and result
description in two sets $\{ V_{i} \}$ (of operators) and $I$ (of indices)
instead of following the protocol provided by Alice
Bob can test Alice knowledge in much simpler way: {\it calculating,
analysing and finally measuring the observable $X=\sum_{i\in I}V_{i}V_{i}^{\dagger}$ }.
Note, by the way, that if third party ``Eve''
copies the message then she also gets full information about $\phi$.

Now one can ask whether quantum information transfer from Alice to
Bob can reduce significantly the information content about $\phi$.
The answer is positive and is contained in the following protocol.

{\it Symmetric projection convincing protocol .-}
As an ancilla Alice sends Bob another copy of $\phi$ (she can prepare it as she knows
the state) and says that the joint measurement projecting the states of both
original particle and the ancilla onto the symmetric subspace
will certainly give positive result. It is easy to see that the
above protocol satisfies both conditions 1 and 2 which any convincing
test should satisfy. Still Bob can not get more information than
the optimal information about unknown $\phi$
extracted from two copies of it.
It is known that such information is far from full one, nevertheless,
it is strictly more than the information one can extract from one copy.
Indeed, if one consider $\phi$ as being chosen randomly by
some previous preparer (who was further in contact with Alice and Bob)
then one can introduce the {\it fidelity}
of estimation of $\phi$ \cite{Pop94} :
\begin{equation}
f \equiv \int_{\phi} d \phi |\langle \phi| \phi_{est}(\phi)\rangle|^2
\label{fide}
\end{equation}
where integral is calculated over uniform distribution of all
pure qubit states and $\phi_{est}(\phi)$ is the state
estimated under the presence of state $\phi$.
It is known that optimal extraction
of information from one copy (say, before Bob
is given an ancilla) is $f_{1copy}=2/3$,
while in the presence of two copies we have
$f_{2copies}=3/4$.
So in the above protocol (which can be called {\it symmetric
convincing projection (SCP) protocol}  still there is a nontrivial
information transfer about the state $\phi$ from Alice to Bob:
after receiving the test message from Alice
Bob can learn more about it (on average $3/4$ in terms of fidelity)
than if he were given the state alone (resp. $2/3$). If Eve
intercepted the complete message (with the ancilla)
she also gets nonzero knowledge about $\phi$
with fidelity $f_{1copy}=2/3$ instead of
$f_{0copy}=1/2$.
Note that in the case of fully classical protocol
the fidelity is $f_{cl}=1$.
So the above SPC protocol is a
legitimate convincing one, being much more closer to hypothetical
``zero knowledge'' than any fully classical protocol.
But it is still not a ``zero knowledge'' one.

{\it Proof of nonexistence of perfect ``zero knowledge''
protocol .-} Here we shall prove that any
protocol with test message satisfying
condition 1 (and hence any convincing Alice message)
has to carry nontrivial information about Bob state $\phi$.
There is even more than that. As we shall see basing on
message satisfying condition 1 Bob (or Eve) can reproduce
with some nonzero probability unknown state from the ancilla.
Suppose that Alice sends Bob the ancilla
in, in general,  mixed state $\varrho$ defined
on the Hilbert space ${\cal H}_{ancilla}$
of arbitrary (may be infinite) dimension,
the classical description of quantum operation $\{ \tilde{V}_{i} \}$
(which, in general, is to be carried out on both the Bob qubit
and the ancilla), and the set of indices $I$ corresponding to
the result $i_{0} \in I$.
The mixed state $\varrho$ describes the ancilla which can
be hydrogen atom, photon with a given state
of its polarisation, molecule with the state
of nuclear spin prepared etc.
The Bob's qubit can be defined as a pure state
of spin of spin-half particle, state of effectively two
level atom and so on. So the model is completely
general from the physical point of view.

After similar considerations as in the case of fully classical
protocols it can be seen that condition 1
is satisfied if and only if the mean values of the following observable
$A=\sum_{i\in I} \tilde{V}_{i}^{\dagger} \tilde{V}_{i}$
(built on the basis of Alice classical part of message) satisfy:
\begin{equation}
 Tr(A \varrho\otimes |\phi\rangle\langle\phi|)=1,
\label{war1}
\end{equation}
\begin{equation}
Tr(A \varrho\otimes |\phi'\rangle \langle \phi'|)<1
\ \mbox{for}\ \ \mbox{any} \ \ \phi'\neq \phi.
\label{war2}
\end{equation}
Note that, as before, the observable $A$ has to have
eigenvalues from the interval $[0,1]$.
So the condition (\ref{war1}) says that
the joint state $\varrho \otimes |\phi\rangle\langle\phi|$
has eigenvectors belonging to the  (may be degenerated) 
eigensubspace of $A$ corresponding to eigenvalue 1.
Let us denote the projector onto that subspace as $P_{A}$
and the orthogonal projection as $P_{A}^{\perp}$.
They both correspond to the subspaces
in the full Hilbert space ${\cal H}={\cal H}_{ancilla}
\otimes {\cal H}_{qubit}$.
First note that $P_{A}$ can not span the full
${\cal H}$ because then $A$ would be identity
and clearly the condition (\ref{war2}) could not
be satisfied.
Hence  $P_{A}^{\perp}$ is represented by {\it nonzero} set of
its eigenvectors $\{ |\Psi_{k}\rangle \}_{k=1}^{N}$ from the full space
${\cal H}$ where $N$ can be, in general, infinite.
From the general theory of Hilbert spaces
we have $|\Psi_{k}\rangle=W_{k} \otimes I |\Psi_{singlet} \rangle$.
Here we have the operators
$W_{k}:{\cal H}_{qubit} \rightarrow {\cal H}_{ancilla}$
(which can be calculated explicitly from
Schmidt decomposition of the corresponding vectors $\Psi_{k}$)
and the familiar two-qubit singlet state
$|\Psi_{singlet} \rangle =\frac{1}{\sqrt{2}}
(|\uparrow_{\mathop{\bf \hat{z}}}
\downarrow_{\mathop{\bf \hat{z}}} \rangle -
| \downarrow_{\mathop{\bf \hat{z}}} \uparrow_{\mathop{\bf \hat{z}}} \rangle)$.
Now a little algebra leads to the conclusion that condition (\ref{war1})
is equivalent to
\begin{equation}
Tr(W_{k}^{\dagger}\varrho W_{k} | \phi^{\perp} \rangle
\langle \phi^{\perp}|)=0
\label{star}
\end{equation}
for all $k=1, ..., N$.
Here $| \phi^{\perp} \rangle$ represents
the (unique) qubit state orthogonal to $| \phi \rangle$.
To derive (\ref{star}) one uses the identity
(see, for instance, \cite{Gis})
$|\phi^{\perp} \rangle=|\sigma_{y} \phi^*\rangle$ where $\sigma_{y}=
i|\uparrow_{\hat{z}} \rangle \langle\downarrow_{\hat{z}}|
-i| \downarrow_{\hat{z}} \rangle \langle \uparrow_{\hat{z}}|$
is familiar Pauli matrix and star $*$ stands for complex conjugation.
Note that we have $W_{k}^{\dagger}:
{\cal H}_{ancilla} \rightarrow {\cal H}_{qubit}$.
So the hermitian operator $W_{k}^{\dagger}\varrho W_{k}$
with positive spectrum is defined on {\it one qubit} Hilbert space.
For (\ref{star}) is equivalent to (\ref{war1})
at least of $2 \times 2$ matrices $\{ W_{k}^{\dagger}\varrho W_{k} \}$
must not vanish as otherwise (\ref{war2}) could not be satisfied.
So for at least one index $k_{0}$ we have
\begin{equation}
p_{k_{0}}\equiv Tr(W_{k_{0}}^{\dagger}\varrho W_{k_{0}})>0
\label{prob}
\end{equation}
Elementary analysis leads to the conclusion that
the matrix $W_{k_{0}}^{\dagger}\varrho W_{k_{0}}$ must represent
projection onto the vector {\it orthogonal} to $\phi^{\perp}$.
But it means that
\begin{equation}
\frac{W_{k_{0}}^{\dagger}\varrho W_{k_{0}}}{p_{k_{0}}}=|\phi\rangle
\langle \phi|.
\end{equation}
Bob does not know  any of $k_{0}$, $\varrho$,
$\phi$. But he can easily compute all $W_{k}$ inferring
(as it was done above) from $P_{A}$ calculated from the description of  quantum operation which he got from Alice.
Suppose now that Alice knows the state and wants to convince Bob
about it. Bob first performs the measurement of $A$.
If he got positive result (i. e. corresponding to eigenvalue $1$)
then he takes the ancilla (in state $\varrho$)
and performs the quantum measurement operation corresponding to
operators
\begin{equation}
\tilde{W}_{k}\equiv
W_{k}^{\dagger}/\sqrt{Tr(\sum_{k=1}^{N}W_{k}^{\dagger}W_{k})}
\label{mes}
\end{equation}
With nonzero probability (\ref{prob}) $p_{k_{0}}$ he will can get the state
$\frac{W_{k_{0}}^{\dagger}\varrho W_{k_{0}}}{p_{k_{0}}}=|\phi\rangle
\langle \phi|$. So there is nonzero chance that he will synthetize the
second copy of unknown qubit state $\phi$.
We must emphasise that once Bob gets the second copy
of $\phi$ in his lab, he {\it knows} about it, as
the presence of the second copy is guaranteed whenever he gets
positive result of the measurement (\ref{mes}).
Note that if Eve gets the full Alice message
she have a chance to reproduce the Bob's qubit state in the same
manner. So we have proved that any convincing message has to carry
highly nontrivial information about $\phi$.
In the case of SPC protocol this information about $\phi$
can be made secure against Eve attack just by teleporting
the second copy of $\phi$ from Alice to Bob.

If we assume that $\phi$ is supposed  be completely random we
can express the result in terms of fidelities of type (\ref{fide})
\cite{Pop}. To this end consider the hidden variable $\eta$ helping
Alice to choose the concrete form of convincing
message she sends if Bob has state $\phi$.
The message is indexed by $\eta$, $\phi$ and is chosen,
in general with probability $P(\eta,\phi)$.
Note that here, in particular,
we allow Alice decisions to be completely random.
Let $Q(\phi)$ be a uniform distribution on pure qubit states $\phi$
and let $p(\eta,\phi)$ represent the probability
(\ref{prob}) which generally can (indirectly)
depend on $\eta$ and $\phi$.
Now consider first the information gain Eve can get
if she intercept the message. Then she can perform the measurement (\ref{mes}).
If she gets the copy (which can happen with nonzero probability)
then she performs optimal estimation on basis of one copy
which involves choice of random variable $\hat{m}$ - the axis
in threedimensional space (see \cite{Pop} for details).
The fidelity of Eve inference can be calculated to be
\begin{eqnarray}
&&f_{Eve}=f_{0 copy}+ \int d \eta d \phi d \hat{m}
P(\eta,\phi) Q(\phi) p(\eta,\phi) \times \nonumber \\
&&\times (|\langle \phi|\phi_{est}(\phi,\hat{m} \rangle)|^{2} \nonumber
-f_{0copy})=f_{0copy}+\Delta
\end{eqnarray}
The integral over random parameter $\hat{m}$ nullifies possible results of
deliberated Alice actions to unable eavesdropping inference.
After performing the integral we have $\Delta=\int d \eta d \phi
P(\eta,\phi) Q(\phi) p(\eta,\phi)(f_{1copy}-f_{0copy})
\equiv \int d \eta d \phi P(\eta,\phi) F(\eta,\phi)$.
As there always exists strictly positive $p_{k_{0}}$
the function $p(\eta,\phi)$ is strictly positive on the whole
probability space, hence $\Delta$ is an integral on the function
$F$ positive everywhere. Thus $\Delta>0$
and
\begin{equation}
f_{Eve}>f_{0copy}.
\end{equation}
So we have proved formally in terms of fidelities
that Alice her message
necessarily carries the information about Bob's state
$\phi$.
The similar analysis can be performed to show that $f_{Bob}>f_{1copy}$.


Finally it is worth to note that Alice is supposed
to convince Bob of her {\it classical knowledge about quantum state}.
The tests of that kind can be of practical significance in future.
If quantum computers eventually are constructed
the question of knowledge about quantum databases
content will probably be important from the point of view
of data security (for example to test whether someone
could have created given data which are under investigation).

On the other hand the analysis of problems of the above
kind provides us new features of interrelation of classical
and quantum information: as we have seen
Alice sends classical text, as well as some quantum state,
so the above model problem satisfies the paradigm where classical
and quantum levels of information are, in general, supposed
to "interact" \cite{tata}. Some implications of the present result
concerning nature of quantum information will be considered
elsewhere. From practical point of view it would be
interesting to consider the result in context of
quantum computing involving single pure quantum bit
\cite{Laf,Park}.

{\it Acknowledgements .-} Most of this work was done when
the authors visited T. J. Watson Research Centre (IBM)
in Yorktown Heights. We thank  Barbara Terhal and John Smolin
for critical comments. Special thanks are due to Charles Bennett
for inspiring discussion and David DiVincenzo for
his deep remarks. The work is partially supported by
Polish Committee for Scientific Research and by European Community
under the grant EQUIP. The work is also supported by John Templeton Foundation. 

{\it Note added.}  In arXiv:1706.06963 E. Adlam and A. Kent have extended our result by providing a thorough quantitative analysis and including 
to relativistic setup. 

%

\end{document}